\documentstyle[12pt,ringberg,amssymb,epsfig]{article}

\begin{document}
\begin{flushright}
\normalsize
\begin{tabular}{r}
UWThPh-1998-47\\
TUM-HEP-322/98\\
SFB-375-303\\
DFTT 43/98\\
hep-ph/9807568
\end{tabular}
\end{flushright}
\vspace{1cm}
\begin{center}
\normalsize\bf
NEUTRINO MASS SPECTRUM AND MIXING
\\
\normalsize\bf
FROM NEUTRINO OSCILLATION DATA\footnote{Talk presented by S.M. Bilenky at the
Ringberg Euroconference
\textit{New Trends in Neutrino Physics},
24--29 May 1998, Ringberg Castle, Tegernsee, Germany.}
\end{center}
\baselineskip=22pt
\centerline{\footnotesize S.M. Bilenky}
\baselineskip=13pt
\centerline{\footnotesize\it Joint Institute for Nuclear Research, Dubna, Russia, and}
\baselineskip=12pt
\centerline{\footnotesize\it Institut f\"ur Theoretische Physik,
Technische Universit\"at Munchen, D--85748 Garching, Germany}
\vspace*{0.3cm}
\centerline{\footnotesize C. Giunti}
\baselineskip=13pt
\centerline{\footnotesize\it INFN, Sezione di Torino, and Dipartimento di Fisica Teorica,
Universit\`a di Torino,}
\baselineskip=12pt
\centerline{\footnotesize\it Via P. Giuria 1, I--10125 Torino, Italy}
\vspace*{0.3cm}
\centerline{\footnotesize W. Grimus}
\baselineskip=13pt
\centerline{\footnotesize\it Institute for Theoretical Physics, University of Vienna,}
\baselineskip=12pt
\centerline{\footnotesize\it Boltzmanngasse 5, A--1090 Vienna, Austria}

\vspace*{0.9cm}
\abstracts{Two schemes of mixing of four massive neutrinos with two close neutrino
masses separated by a gap of
$\Delta{m}^2 \sim 1 \, {\rm eV}^2$
can accommodate
solar, atmospheric and LSND
neutrino oscillation data.
It is shown that long-baseline $\bar\nu_e\to\bar\nu_x$ and $\nu_\mu\to\nu_e$
transitions are strongly suppressed in these
schemes. The scheme of mixing of three massive neutrinos with a mass
hierarchy that can describe solar and atmospheric data is also discussed.
It is shown that in this scheme the effective Majorana mass that
characterizes the matrix
element of neutrinoless double-beta decay is smaller than
$\sim 10^{-2} \, {\rm eV}$.}
 
\normalsize\baselineskip=15pt
\setcounter{footnote}{0}
\renewcommand{\thefootnote}{\alph{footnote}}

\section{Introduction}
\label{Introduction}

The strong evidence in favor of oscillations of atmospheric neutrinos
obtained in the Super-Kamiokande experiment\cite{SK-atm}
made the problem of
neutrino masses and mixing one of the central problem of the
physics of elementary
particles. 
The Super-Kamiokande evidence is the first important step in the
investigation
of the phenomenon of neutrino mixing
proposed many years
ago by B. Pontecorvo\cite{Pontecorvo}.  
There is no doubt that an understanding of the physical origin of neutrino
masses and mixing will require many new experiments.

The Super-Kamiokande data can be explained with
$\nu_\mu\to\nu_\tau$ or $\nu_\mu\to\nu_s$
oscillations with
\begin{equation}
5 \times 10^{-4} \, {\rm eV}^2
\lesssim
\Delta{m}^2_{{\rm atm}}
\lesssim
5 \times 10^{-3} \, {\rm eV}^2
\label{atm}
\end{equation}
and a large mixing angle.

Indications in favor of neutrino mixing were obtained also in solar
neutrino experiments.
From the analysis of the existing data it follows
that\cite{SK-sun}
\begin{equation}
\Delta{m}^2_{{\rm sun}}
\sim
10^{-5} \, {\rm eV}^2
\,
\mbox{(MSW)}
\quad
\mbox{or}
\quad
\Delta{m}^2_{{\rm sun}}
\sim
10^{-10} \, {\rm eV}^2
\,
\mbox{(vac. osc.)}
\,.
\label{sun}
\end{equation}

Finally,
indications in favor of $\bar\nu_\mu\to\bar\nu_e$ oscillations were
obtained in the accelerator LSND experiment\cite{LSND}.
If all other data on the search for $\nu_\mu\to\nu_e$ transitions
in short-baseline (SBL) experiments are taken
into account from the analysis of the data of this experiment it follows
that
\begin{equation}
0.3 \, {\rm eV}^2
\lesssim
\Delta{m}^2_{{\rm SBL}}
\lesssim
2.2 \, {\rm eV}^2
\,.
\label{SBL}
\end{equation}

We will discuss here what conclusion about
the neutrino mass spectrum and
the elements of neutrino mixing matrix can be obtained from the results of all
neutrino oscillation experiments. We will consider also some consequences
for the future experiments that can be inferred from the model
independent analysis
of the existing data.

We will present in the beginning the general theoretical framework
of neutrino mixing\cite{Bilenky}.

\section{Phenomenological theory of neutrino mixing}
\label{Phenomenological theory of neutrino mixing}

All existing data on the investigation of neutrino processes 
are perfectly described by the standard
charged-current (CC) and neutral-current (NC) Lagrangians
\begin{eqnarray}
&&
{\cal L}_I^{{\rm CC}}
=
- \frac{g}{\sqrt{2}}
\sum_{\ell=e,\mu,\tau}
\overline{\nu_{{\ell}L}} \, \gamma_{\alpha} \, \ell_L
\, W^{\alpha}
+
\mbox{h.c.}
\,,
\label{LCC}
\\
&&
{\cal L}_I^{{\rm NC}}
=
- \frac{g}{2\,\cos\theta_W}
\sum_{\ell=e,\mu,\tau}
\overline{\nu_{{\ell}L}} \, \gamma_{\alpha} \, \nu_{{\ell}L}
\, Z^{\alpha}
+
\mbox{h.c.}
\,.
\label{LNC}
\end{eqnarray}
The CC and NC interaction Lagrangians (\ref{LCC}) and (\ref{LNC}) conserve electron
$L_e$,
muon $L_\mu$ and tau $L_\tau$
lepton numbers and CC interactions determine the
notion of flavor neutrinos $\nu_\ell$ and antineutrinos $\bar\nu_\ell$
($\ell=e,\mu,\tau$). 
There are no indications in favor of violation of lepton
numbers in weak processes and from the existing experiments very strong
bounds on the relative probabilities $R$ of lepton number violating
processes were obtained. For example, for 
$ \mu \to e \, \gamma $,
$ \mu \to 3 \, e $
it was found: 
\begin{equation}
R_{\mu \to e \, \gamma}
\leq
5 \times 10^{-11}
\,,
\qquad
R_{\mu \to 3 \, e}
\leq
10^{-12}
\,.
\label{R}
\end{equation}

The neutrino mixing hypothesis\cite{Pontecorvo,Bilenky}
is based on the assumption that
neutrino masses are
different from zero and the neutrino mass term does not conserve lepton
numbers.

Only a Dirac mass term is allowed in the case of
quarks.
This is connected with the fact that quarks are charged particles.
Massive neutrinos can be Dirac
particles (if the total lepton number $L=L_e+L_\mu+L_\tau$ is conserved)
or Majorana
particles (if neutrino mass term does not conserve any lepton number).

The Dirac neutrino mass term has the form
\begin{equation}
{\cal L}^{{\rm D}}
=
-
\sum_{\alpha,\beta}
\overline{\nu_{{\alpha}R}}
\,
M^{{\rm D}}_{\alpha\beta}
\,
\nu_{{\beta}L}
+
\mbox{h.c.}
\,,
\label{dirac}
\end{equation}
where $M^{{\rm D}}$ is a complex $3\times3$ non-diagonal matrix.
It is obvious that the Dirac mass term conserves total lepton number.
After the standard
diagonalization,
for the left-handed fields
$\nu_{{\alpha}L}$
we have
\begin{equation}
\nu_{{\alpha}L}
=
\sum_{i=1}^{3}
U_{{\alpha}i} \, \nu_{iL}
\label{mix}
\end{equation}
where $\nu_i$ is the field of the Dirac neutrino with mass $m_i$ ($i=1,2,3$)
and $U$ is the unitary mixing matrix.
The Dirac mass term can be generated by the same standard Higgs
mechanism with which the masses of quarks
and
leptons are generated.

The general Majorana mass term that does not conserve lepton numbers
has the form
\begin{equation}
{\cal L}^{{\rm M}}
=
{\cal L}^{{\rm M}}_{L}
+
{\cal L}^{{\rm D}}
+
{\cal L}^{{\rm M}}_{R}
\,.
\label{majorana1}
\end{equation}
with
\begin{equation}
{\cal L}^{{\rm M}}_{L}
=
- \frac{1}{2}
\sum_{\alpha,\beta}
\overline{(\nu_{{\alpha}L})^c}
\,
M^{{\rm L}}_{\alpha\beta}
\,
\nu_{{\beta}L}
+
\mbox{h.c.}
\,.
\label{majorana2}
\end{equation}
Here $M^{{\rm L}}$
is a complex $3\times3$ symmetric matrix and
$ (\nu_{{\alpha}L})^c \equiv {\cal C} \overline{\nu_{{\alpha}L}}^T $
(${\rm C}$ is the charge conjugation matrix).
The mass term
${\cal L}^{{\rm M}}_{R}$
can be obtained from Eq.(\ref{majorana2}) with the change
$L \to R$.

After the 
diagonalization of the mass term (\ref{majorana1})
we have
\begin{equation}
\nu_{{\alpha}L}
=
\sum_{i=1}^{n}
U_{{\alpha}i} \, \nu_{iL}
\,,
\quad
(\nu_{aR})^c
=
\sum_{i=1}^{n}
U_{ai} \, \nu_{iL}
\,.
\label{02}
\end{equation} 
where
$ \nu_i = \nu_i^c $
is a Majorana field with mass $m_i$. 

The Majorana mass term (\ref{majorana1})
can be generated only in the framework of theories beyond the Standard
Model.
In this mass term  three flavor left-handed fields
$\nu_{{\alpha}L}$
and
three right-handed fields
$\nu_{{\alpha}R}$
enter. The number of massive Majorana
particles is equal in this case to six.
In general,
if the number of right-handed fields that enter into
the mass term is equal to $n_R$,
the number
of massive Majorana fields is equal to
$ n = 3 + n_R $.

Let us notice that in the case of a left-handed Majorana mass term
\begin{equation}
{\cal L}^{{\rm M}}
=
{\cal L}^{{\rm M}}_{L}
\label{majorana}
\end{equation}
only flavor left-handed fields enter into Lagrangian. The number of
Majorana neutrinos is equal in this case to three.

Two possible options are usually discussed in the Majorana case:

\begin{enumerate}

\item
The \emph{ see-saw} option\cite{see-saw}.

If the lepton numbers are violated by the right-handed Majorana mass term
at a mass scale much larger than the electroweak scale,
the Majorana neutrino mass spectrum is composed by
three light masses $m_i$ ($i=1,2,3$)
and three very heavy masses $M_i$ ($i=1,2,3$)
that characterize the scale of lepton number violation.
The light neutrino masses are given in this case by the see-saw formula
\begin{equation}
m_i \sim \frac{ ( m_i^F )^2 }{ M_i }
\ll
m_i^F
\quad
(i=1,2,3)
\,,
\label{011}
\end{equation}
where $m_i^F$ is the mass of
the charged lepton or up-quark in the $i^{{\rm th}}$ generation.
The see-saw mechanism provides a
plausible explanation for the smallness of neutrino masses
with respect to the masses of all other fundamental fermions.

\item
The \emph{sterile neutrino} option.

If all the Majorana masses in Eq.(\ref{majorana1})
are small,
active neutrinos
$\nu_e$, $\nu_\mu$ and $\nu_\tau$
can transfer into the sterile particles $\nu_{as}$
that are quanta of the right-handed fields
$\nu_{aR}$.
Notice that  sterile neutrinos can appear in the framework of
see-saw
mechanism under some additional assumptions
(``singular see-saw''\cite{CKL98}).

\end{enumerate}

We will consider two possible scenarios:

\begin{enumerate}

\item
All three indications in favor of neutrino
oscillations are confirmed.

\item
Only the solar and atmospheric neutrino indications in favor of neutrino
mixing
are confirmed.

\end{enumerate}

\begin{table}[t!]
\begin{tabular*}{\linewidth}{@{\extracolsep{\fill}}cc}
\begin{minipage}{0.47\linewidth}
\begin{center}
\setlength{\unitlength}{1cm}
\begin{tabular*}{0.95\linewidth}{rlrl}
I:
&
\raisebox{-2cm}{
\begin{picture}(1.5,4)
\thicklines
\put(0.5,0.2){\vector(0,1){3.8}}
\put(0.4,0.2){\line(1,0){0.2}}
\put(0.8,0.15){\makebox(0,0)[l]{$m_1$}}
\put(0.4,0.4){\line(1,0){0.2}}
\put(0.8,0.45){\makebox(0,0)[l]{$m_2$}}
\put(0.4,0.8){\line(1,0){0.2}}
\put(0.8,0.8){\makebox(0,0)[l]{$m_3$}}
\put(0.4,3.5){\line(1,0){0.2}}
\put(0.8,3.5){\makebox(0,0)[l]{$m_4$}}
\end{picture}
}
&
II:
&
\raisebox{-2cm}{
\begin{picture}(1.5,4)
\thicklines
\put(0.5,0.2){\vector(0,1){3.8}}
\put(0.4,0.2){\line(1,0){0.2}}
\put(0.8,0.2){\makebox(0,0)[l]{$m_1$}}
\put(0.4,2.9){\line(1,0){0.2}}
\put(0.8,2.9){\makebox(0,0)[l]{$m_2$}}
\put(0.4,3.3){\line(1,0){0.2}}
\put(0.8,3.25){\makebox(0,0)[l]{$m_3$}}
\put(0.4,3.5){\line(1,0){0.2}}
\put(0.8,3.55){\makebox(0,0)[l]{$m_4$}}
\end{picture}
}
\\
&&&
\\
&&&
\\
IIIA:
&
\raisebox{-2cm}{
\begin{picture}(1.5,4)
\thicklines
\put(0.5,0.2){\vector(0,1){3.8}}
\put(0.4,0.2){\line(1,0){0.2}}
\put(0.8,0.2){\makebox(0,0)[l]{$m_1$}}
\put(0.4,0.6){\line(1,0){0.2}}
\put(0.8,0.6){\makebox(0,0)[l]{$m_2$}}
\put(0.4,3.3){\line(1,0){0.2}}
\put(0.8,3.25){\makebox(0,0)[l]{$m_3$}}
\put(0.4,3.5){\line(1,0){0.2}}
\put(0.8,3.55){\makebox(0,0)[l]{$m_4$}}
\end{picture}
}
&
IIIB:
&
\raisebox{-2cm}{
\begin{picture}(1.5,4)
\thicklines
\put(0.5,0.2){\vector(0,1){3.8}}
\put(0.4,0.2){\line(1,0){0.2}}
\put(0.8,0.15){\makebox(0,0)[l]{$m_1$}}
\put(0.4,0.4){\line(1,0){0.2}}
\put(0.8,0.45){\makebox(0,0)[l]{$m_2$}}
\put(0.4,3.1){\line(1,0){0.2}}
\put(0.8,3.1){\makebox(0,0)[l]{$m_3$}}
\put(0.4,3.5){\line(1,0){0.2}}
\put(0.8,3.5){\makebox(0,0)[l]{$m_4$}}
\end{picture}
}
\end{tabular*}
\end{center}
\end{minipage}
&
\begin{minipage}{0.47\linewidth}
\begin{center}
\mbox{\epsfig{file=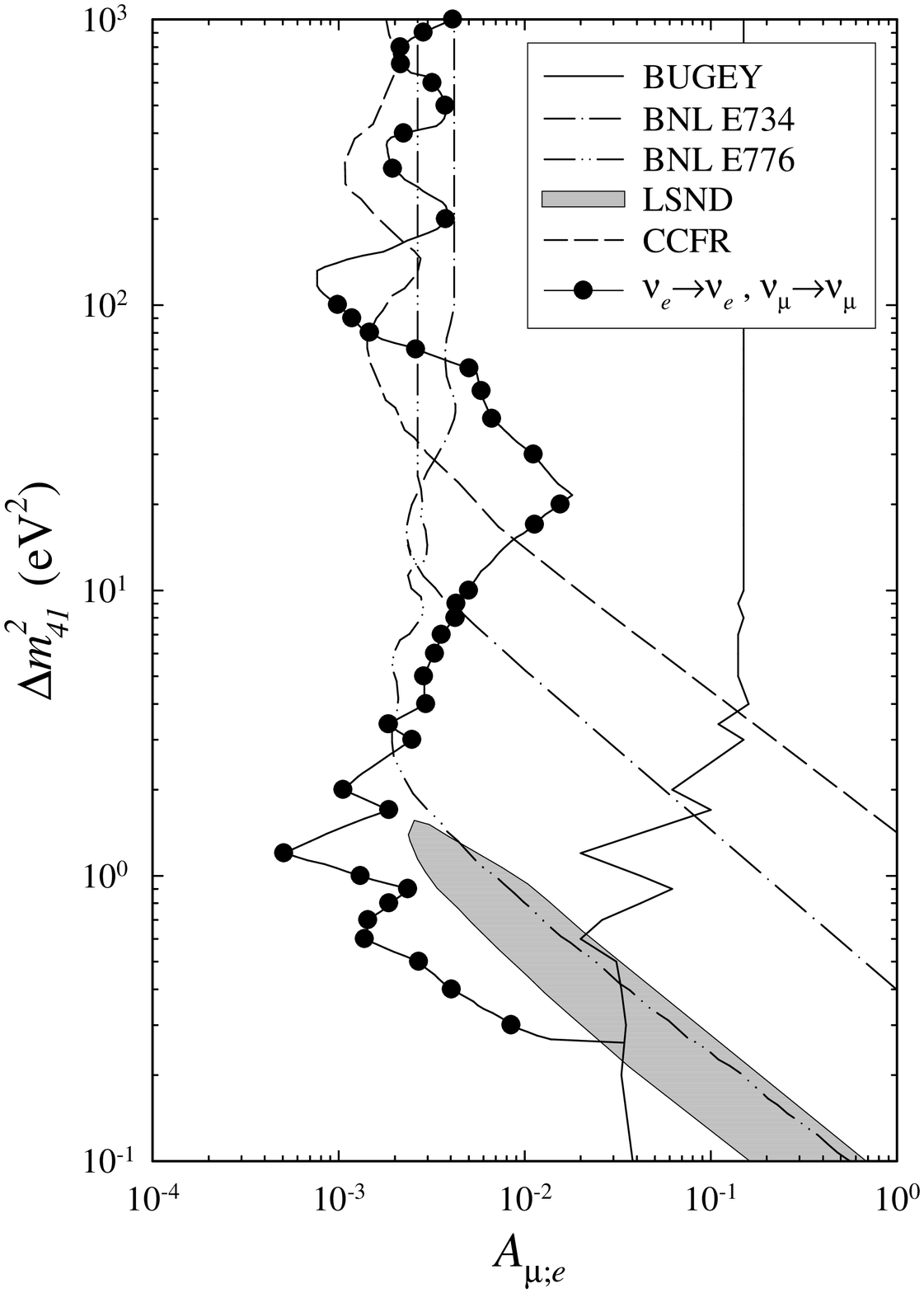,width=0.95\linewidth}}
\end{center}
\end{minipage}
\\
\refstepcounter{figure}
\label{fig1}                 
Figure \ref{fig1}
&
\refstepcounter{figure}
\label{fig2}                 
Figure \ref{fig2}
\end{tabular*}
\null \vspace{-0.5cm} \null
\end{table}

\section{Four massive neutrinos}
\label{Four massive neutrinos}

At least four massive neutrinos are needed\cite{four,BGKP,BGG96,BGG97a,BGG97-98,CKL98}
in order
to have three different scales of
$\Delta{m}^2$.
The three types of neutrino mass spectra that can accommodate 
the solar, atmospheric and LSND scales of $\Delta{m}^2$
are shown in Fig.\ref{fig1}.
In all these mass spectra
there are two groups of close masses
separated by a gap of the order of 1 eV
which gives
$
\Delta{m}^2_{41} \equiv m_4^2 - m_1^2
\simeq \Delta{m}^2_{{\rm LSND}}
\sim
1 \, {\rm eV}^2
$.

Only the
largest mass-squared difference
$ \Delta{m}^2_{41} $
is relevant
for the oscillations in short-baseline (SBL) experiments
and the SBL transition probabilities have the same
dependence on the parameter
$ \Delta{m}^2_{41} L / 2 p $
as the standard
two-neutrino probabilities\cite{BGKP}:
\begin{eqnarray}
&&
P_{\nu_\alpha\to\nu_\beta}
=
\frac{1}{2} \, A_{\alpha;\beta} \,
\left( 1 - \cos \frac{ \Delta{m}^2_{41} L }{ 2 p } \right)
\,,
\label{Ptran}
\\
&&
P_{\nu_\alpha\to\nu_\alpha}
=
1
-
\frac{1}{2} \, B_{\alpha;\alpha} \,
\left( 1 - \cos \frac{ \Delta{m}^2_{41} L }{ 2 p } \right)
\,.
\label{Psurv}
\end{eqnarray}
Here $L$ is
the source-detector distance
and $p$ is the neutrino momentum.

The oscillation amplitudes
$A_{\alpha;\beta}$ and $B_{\alpha;\alpha}$
depend on the elements on the mixing matrix $U$
and on the form of the neutrino mass spectrum:
\begin{eqnarray}
&&
A_{\alpha;\beta}
=
4
\left|
\sum_i
U_{{\beta}i}
\,
U_{{\alpha}i}^*
\right|^2
\,,
\label{Aab}
\\
&&
B_{\alpha;\alpha}
=
4
\left( \sum_i |U_{{\alpha}i}|^2 \right)
\left( 1 - \sum_i |U_{{\alpha}i}|^2 \right)
\,,
\label{Baa}
\end{eqnarray}
where the index $i$ runs over the indices of the first or
(because of the unitarity of $U$) second group
of neutrino masses.

The results of SBL reactor
$\bar\nu_e$
and
accelerator
$\nu_\mu$
disappearance
experiments in which no oscillations were found
imply that
$ B_{\alpha;\alpha} \leq B_{\alpha;\alpha}^0 $
for
$\alpha=e,\mu$.
The upper bounds
$B_{\alpha;\alpha}^0$
for the amplitudes
$B_{\alpha;\alpha}$
are given by the exclusion curves of 
SBL disappearance experiments
and depend on the value of $\Delta{m}^2_{41}$.
Using Eq.(\ref{Baa}),
these upper bounds
imply
the following constraints for
the quantities
$ \sum_i |U_{{\alpha}i}|^2 $
($\alpha=e,\mu$):
\begin{equation}
\sum_i |U_{{\alpha}i}|^2 \leq a_\alpha^0
\quad \mbox{or} \quad
\sum_i |U_{{\alpha}i}|^2 \geq 1 - a_\alpha^0
\,,
\label{05}
\end{equation}
where
\begin{equation}
a_{\alpha}^0
=
\frac{1}{2}
\left( 1 - \sqrt{ 1 - B_{\alpha;\alpha}^0 } \,\right)
\,.
\label{06}
\end{equation}
The most stringent values of
$a_e^0$ and $a_\mu^0$
are given by the results of the
the Bugey reactor experiment\cite{Bugey95}
and the CDHS\cite{CDHS84} and CCFR\cite{CCFR84}
accelerator experiments.

We have considered
the range
$
10^{-1} \leq
\Delta{m}^2_{41}
\leq 10^3 \, {\rm eV}^2
$.
In this range
$ a_e^0 \lesssim 4 \times10^{-2} $
and
$ a_\mu^0 \lesssim 2 \times 10^{-1} $
for $ \Delta{m}^2_{41} \gtrsim 0.3 \, {\rm eV}^2 $.
Thus, from the results of disappearance experiments
it follows that
$ \sum_i |U_{ei}|^2 $
and
$ \sum_i |U_{{\mu}i}|^2 $
can be either small or large (close to one).

From the four possibilities for the quantities
$ \sum_i |U_{ei}|^2 $
and
$ \sum_i |U_{{\mu}i}|^2 $
(small-small, small-large, large-small and large-large)
for each neutrino mass spectrum in Fig.1 only one possibility is
compatible with the results
of solar and atmospheric neutrino
experiments\cite{BGKP,BGG96}.

In the case of spectra I and II we have
\begin{equation}
|U_{ek}|^2 \leq a_e^0
\qquad \mbox{and} \qquad
|U_{{\mu}k}|^2 \leq a_\mu^0
\,,
\label{07}
\end{equation}
with $k=4$ for the mass spectrum I and $k=1$ for the mass spectrum II.
In the case of spectrum IIIA we have
\begin{equation}
\sum_{i=1,2} |U_{ei}|^2 \leq a_e^0
\qquad \mbox{and} \qquad
\sum_{i=1,2} |U_{{\mu}i}|^2 \geq 1 - a_\mu^0
\,,
\label{08}
\end{equation}
whereas in the case of spectrum IIIB we have
\begin{equation}
\sum_{i=3,4} |U_{ei}|^2 \leq a_e^0
\qquad \mbox{and} \qquad
\sum_{i=3,4} |U_{{\mu}i}|^2 \geq 1 - a_\mu^0
\,.
\label{09}
\end{equation}

In the case of spectra I and II 
$\nu_\mu\to\nu_e$
transitions in SBL experiments are strongly suppressed.
In fact the upper bound of $ A_{e\mu} $ is given by

\begin{equation}
A_{e;\mu}
\leq
4 \, |U_{ek}|^2 \, |U_{{\mu}k}|^2
\leq
4 \, a_e^0 \, a_\mu^0
\,.
\label{10}
\end{equation}
In Fig.\ref{fig2} the upper bound
(\ref{10}) is compared with the latest LSND-allowed region (90\% CL).
Fig.\ref{fig2} shows that
the spectra of type I and II
(that include also the hierarchical spectrum)
are disfavored by the result of the LSND
experiment
(they are compatible with the results of the LSND experiment
only in the narrow region of $\Delta{m}^2_{41}$ around
$ 0.2 - 0.3 \, {\rm eV}^2 $,
where there is no information
on $B_{\mu;\mu}$).
On the other hand, 
it is easy to show that 
spectra IIIA and IIIB are compatible with the results of the LSND
experiment.
Thus we come to the conclusion that from all possible spectra of four
massive
neutrinos only spectra IIIA and IIIB are favored
by the data of LSND and all other neutrino oscillation experiments.

\begin{table}[t!]
\begin{tabular*}{\linewidth}{@{\extracolsep{\fill}}cc}
\begin{minipage}{0.47\linewidth}
\begin{center}
\mbox{\epsfig{file=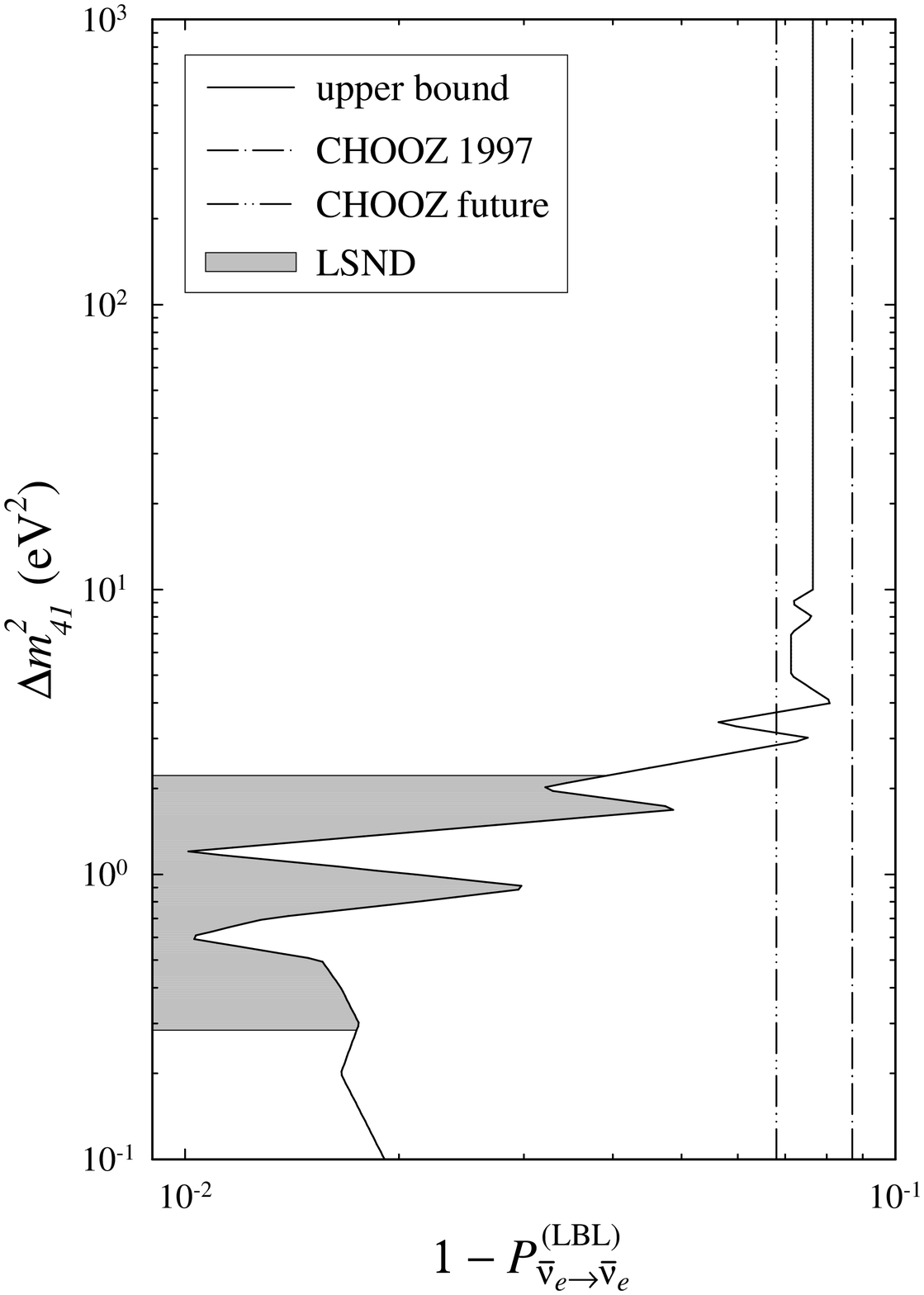,width=0.95\linewidth}}
\end{center}
\end{minipage}
&
\begin{minipage}{0.47\linewidth}
\begin{center}
\mbox{\epsfig{file=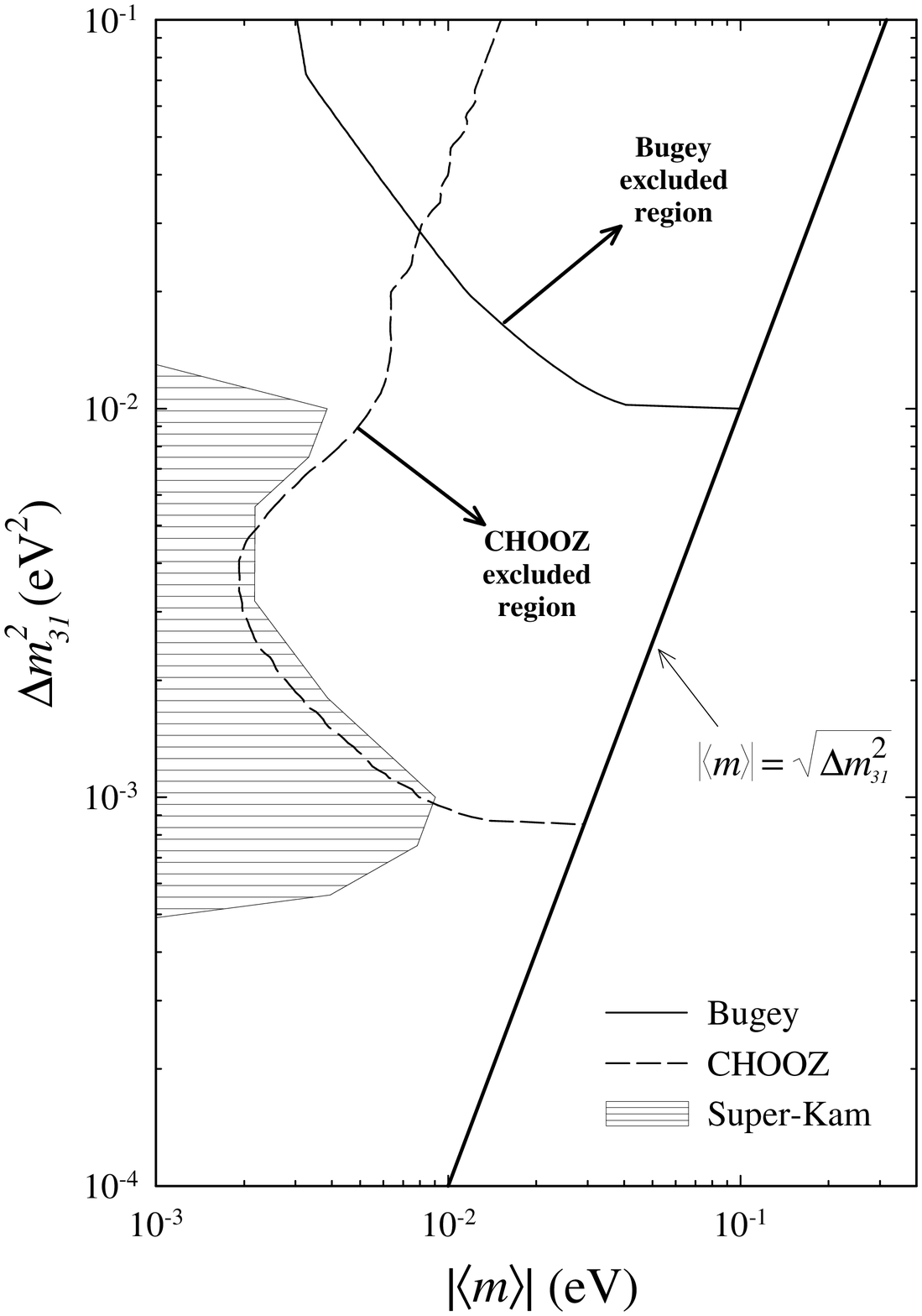,width=0.95\linewidth}}
\end{center}
\end{minipage}
\\
\refstepcounter{figure}
\label{fig3}                 
Figure \ref{fig3}
&
\refstepcounter{figure}
\label{fig4}                 
Figure \ref{fig4}
\end{tabular*}
\null \vspace{-0.5cm} \null
\end{table}

We discuss now some consequences of the schemes with mass spectra IIIA and
IIIB
for future experiments.
Let us consider in the framework of these two schemes
the value of the effective mass 
$m(^3{\rm H})$ measured in
tritium $\beta$-decay experiments and the value 
of the effective Majorana mass 
$
|\langle{m}\rangle|
\equiv
\left| \sum_k U_{ek}^2 \, m_k \right|
$
measured in
neutrinoless double-$\beta$ decay experiments.
In the schemes IIIA and IIIB we have respectively
\begin{equation}
m(^3{\rm H})
\simeq
m_4
\,,
\quad
|\langle{m}\rangle|
\leq m_4
\,,
\label{14}
\end{equation}
and
\begin{equation}
\begin{array}{l} \displaystyle
m(^3{\rm H})
\leq
a_e^0 \, m_4 \ll m_4
\,,
\\ \displaystyle
|\langle{m}\rangle|
\leq
a_e^0 \, m_4 \ll m_4
\,.
\end{array}
\label{15}
\end{equation}
Therefore,
if the scheme IIIA is realized in nature,
there is a possibility to see the effects of the relatively large
neutrino mass  $m_4 \simeq \sqrt{\Delta{m}^2_{41}}$
in future tritium $\beta$-decay experiments
 and
in neutrinoless double-$\beta$ decay experiments.

Let us consider now neutrino transitions in long-baseline (LBL) experiments.
In the scheme IIIA the LBL transition probabilities are given by\cite{BGG97a}
\begin{equation}
P_{\nu_\alpha\to\nu_\beta}^{{\rm LBL}}
=
\left|
\sum_{k=1,2} U_{{\alpha}k}^* \, e^{-i\frac{\Delta{m}^2_{k1}L}{2E}} \, U_{{\beta}k}
\right|^2
+
\left| \sum_{j=3,4} U_{{\alpha}j}^* \, U_{{\beta}j} \right|^2
\,.
\label{16}
\end{equation}
The transition probabilities in the scheme IIIB can be obtained from
(\ref{16})
with the change $1,2\leftrightarrows3,4$.
The inequalities 
(\ref{08}) and (\ref{09})
imply rather strong constraints
on the probabilities of
$\bar\nu_e\to\bar\nu_e$
and
$\nu_\mu\to\nu_e$
transitions in LBL experiments\cite{BGG97a}.
Indeed, for the probability of $\bar\nu_e\to\bar\nu_e$ transitions
we have
\begin{equation}
P_{\bar\nu_e\to\bar\nu_e}^{{\rm LBL}}
\geq
\left( \sum_{j=3,4} |U_{ej}|^2 \right)^2
\geq
(1-a_e^0)^2
\label{17}
\end{equation}
in scheme IIIA and
\begin{equation}
P_{\bar\nu_e\to\bar\nu_e}^{{\rm LBL}}
\geq
\left( \sum_{k=1,2} |U_{ej}|^2 \right)^2
\geq
(1-a_e^0)^2
\label{18}
\end{equation}
in scheme IIIB.
Hence,
in both schemes IIIA and IIIB
$P_{\bar\nu_e\to\bar\nu_e}^{{\rm LBL}}$
is close to one and we expect that the LBL transition probability of 
$\bar\nu_e$
into any other state is small.
Indeed,
in both schemes we have
\begin{equation}
1 - P_{\bar\nu_e\to\bar\nu_e}^{{\rm LBL}}
\leq
a_e^0 \, (2-a_e^0)
\,.
\label{19}
\end{equation}
This limit is shown by the solid line in Fig.\ref{fig3}.
The exclusion line obtained in the 
CHOOZ experiment\cite{CHOOZ}
(dash-dotted line)
and the final sensitivity of the CHOOZ experiment
(dash-dot-dotted line) are also shown.
It can be seen that for
$ \Delta{m}^2_{41} \lesssim 1 \, {\rm eV}^2 $
the upper bound (\ref{19}) for
$ 1 - P_{\bar\nu_e\to\bar\nu_e}^{{\rm LBL}} $
is much smaller than
the upper bound reached in CHOOZ experiment
and also much smaller than the final sensitivity of the CHOOZ experiment.

\section{Three massive neutrinos}
\label{Three massive neutrinos}

If the results of the LSND experiment
will not be confirmed by future experiments,
the most plausible scheme is the one with mixing of three
massive
neutrinos and a mass hierarchy\cite{three,BBGK}:
\begin{equation}
m_1 \ll m_2 \ll m_3
\,.
\label{20}
\end{equation}

The investigations of
neutrino oscillations does not allow\cite{Bilenky}
to answer the fundamental question:
are massive neutrinos Dirac or Majorana particles?
Only investigations of neutrinoless double-$\beta$ decay could allow to
answer this question.
In the case of a three-neutrino
mass hierarchy for the effective Majorana mass we have\cite{BGKM}
\begin{equation}
|\langle{m}\rangle|
\simeq
|U_{e3}|^2 \, \sqrt{\Delta{m}^2_{31}}
\,.
\label{21}
\end{equation}

The results of reactor neutrino experiments
imply\cite{BBGK} an upper bound for
$|U_{e3}|^2$:
$ |U_{e3}| \leq a_e^0 $,
with $a_e^0$ given in Eq.(\ref{06}).
Therefore the effective Majorana mass
is bounded by
\begin{equation}
|\langle{m}\rangle|
\lesssim
a_e^0 \, \sqrt{\Delta{m}^2_{31}}
\,.
\label{22}
\end{equation}
The value of this upper bound as a function $\Delta{m}^2_{31}$
obtained from 90\% CL
exclusion plots of the Bugey\cite{Bugey95} and CHOOZ\cite{CHOOZ}
experiments
is presented in Fig.\ref{fig4}
(the solid and dashed line, respectively).
The region on the right of the thick straight solid line
is forbidden by the unitarity bound
$
|\langle{m}\rangle|
\leq
\sqrt{\Delta{m}^2_{31}}
$.

Also the results of
the Super-Kamiokande atmospheric neutrino experiment\cite{SK-atm}
imply an upper bound for
$|U_{e3}|^2$.
The shadowed region in Fig.\ref{fig4}
shows the
region allowed by Super-Kamiokande results at 90\% CL
that we have obtained
using the results of three-neutrino analysis performed by Yasuda\cite{Yasuda}.

Figure \ref{fig4} shows that the results of the
Super-Kamiokande and CHOOZ experiments
imply that
$
|\langle{m}\rangle|
\lesssim
10^{-2} \, {\rm eV}
$.

The observation of neutrinoless double-$\beta$
decay with a probability
that corresponds to a value of
$|\langle{m}\rangle|$
significantly larger than
$10^{-2} \, {\rm eV}$
would mean that
the masses of three neutrinos do not have a hierarchical pattern
and/or exotic mechanisms (right-handed currents, supersymmetry
with violation of R-parity, \ldots\cite{exotic})
are responsible for the process.

Let us notice that from the results of the Heidelberg-Moscow
$^{76}$Ge experiment\cite{Heidelberg-Moscow}
it follows that
$
|\langle{m}\rangle|
\lesssim
0.5 - 1.5 \, {\rm eV}
$.
The next generation\cite{next0bb} of neutrinoless double-$\beta$ experiments
will reach
$
|\langle{m}\rangle|
\simeq
10^{-1} \, {\rm eV}
$.
Possibilities to reach
$
|\langle{m}\rangle|
\simeq
10^{-2} \, {\rm eV}
$
are under discussion\cite{next0bb}.

\section{Conclusions}
\label{Conclusions}

The neutrino mass spectrum and the structure of the neutrino
mixing matrix depend crucially on the confirmation
of the results of the LSND experiment.
If this results will be confirmed
we need (at least) four massive neutrinos 
with mass spectrum  of type IIIA or IIIB
(see Fig.\ref{fig1}).
If the results of the LSND experiment will not be confirmed,
a plausible scenario is the one with
three massive neutrinos and a mass hierarchy.
The investigation of the nature of massive neutrinos
(Dirac or Majorana?)
will require in this case
to reach a sensitivity at the level of
$ 10^{-2} \, {\rm eV} $
in the search for
neutrinoless double-$\beta$ decay. 

\bigskip

S.M.B. acknowledge
the support of the ``Sonderforschungsbereich 375-95 fuer
Astro-Teilchenphysik der Deutschen Forschungsgemeinschaft''.

\end{document}